\documentclass[10pt]{iopart}
\input epsf

\newcommand{\beq}{\begin{equation}}
\newcommand{\eeq}[1]{\label{#1} \end{equation}}
\newcommand{\beqar}{\begin{eqnarray}}
\newcommand{\eeqar}[1]{\label{#1} \end{eqnarray}}

\begin{document}

\title[A Microscopic Picture of Strangeness Production at the SPS]{}

\author{Stephen E. Vance \footnote[3]{vance@bnl.gov}}

\address{Brookhaven National Laboratory, Upton, NY 11973}

%OK, could be better.  Shorten sentences and give more punch. 

% Completely rewrite
\begin{abstract}

The microscopic origin of the strangeness enhancement observed in PbPb
collisions at the SPS is studied by testing the effects of three
types of strangeness enhancement mechanisms. 
First, the enhancements from the junction stopping and  
junction loop mechanisms found in the HIJING/B\=B event 
generator are reviewed showing that they can account for only part  
of the observed hyperon enhancements.   Including the 
efficient final state chemical changing processes 
(e.g. $\Lambda + K  \leftrightarrow  \Xi  + \pi$)  by coupling the 
General Cascade Program (GCP) to HIJING/B\=B are also only able to 
account for part of the hyperon enhancements.  In addition, it is shown 
that HIJING/B\=B+GCP can not reproduce the enhancement 
of the $K^+/ \pi^+$ ratio.  The inclusion of 
transient field fluctuations or ropes to HIJING/B\=B+GCP is 
therefore shown to be needed to account for the observed enhancements. 
\end{abstract}



\maketitle

\section{Introduction}

Striking enhancements of multistrange baryons have been recently observed 
in PbPb collisions at the SPS.  In fact, when comparing the hyperons produced
per number of participant, a
factor of 20 enhancement is observed for the $\Omega$ in central $PbPb$ 
collisions\cite{wa97_qm99,wa97_plb} over that found
in $pPb$ and $pBe$ collisions.   In addition, the $K^+/ \pi^+$ in central
$PbPb$ collisions is found to be twice the value observed in $pp$ collisions
\cite{na49_qm99}.

These large enhancements are of particular interest 
since they have been proposed as one of the signatures of the 
formation of the quark-gluon plasma\cite{rafelski_82,rafelski_prl82,koch_86}.
Several different approaches which assume a plasma initial 
condition are able to describe the observed yields.  
Different statistical fireball 
models (see \cite{letessier00} and \cite{bmunzinger99}) are successful 
at demonstrating the statistical nature of the ratios of the hadrons.   
The ALCOR model which assumes the formation of a constituent quark plasma 
followed by a coalescence of the quarks into hadrons has also been able 
to successfully describe the hadronic yields and ratios\cite{alcor}.
In addition, a more elaborate model which begins with a system initially 
in a plasma which then freezes out by undergoing a first order phase 
transition followed by the final state rescattering of the produced particles 
is able to describe both the yields and the transverse momentum 
spectra of the hyperons\cite{bass00}.

A different approach is taken in this report, where an initial plasma
state is not assumed and where different microscopic hadronic 
mechanisms are studied.  Here, it is found that three
hadronic mechanisms can provide large 
strangeness enhancements which can effectively describe the data.    
  
The first mechanism is the baryon junction 
stopping\cite{khar96,vanceHijb} and baryon junction pair 
production\cite{vanceHijbb} processes.  
These Regge-motivated processes enhance 
baryon production at mid-rapidity since they produce baryons in the 
mid-rapidity region which are completely composed of sea quarks.
The second is final state interactions or the scattering of the 
newly produced hadrons\cite{vanceHijGcp}.  Here, the most efficient 
interaction is the exothermic, strangeness exchange interaction 
(e.g. $N + K^*  \rightarrow \Lambda + \pi$) which produces a hyperon 
with one greater net strangeness.   
While the above mechanisms enhance the hyperons, they do not describe
the total strangeness enhancement of the system as observed in the 
$K^+/ \pi^+$ ratio.  Thus, a third mechanism is needed.   
Transient fluctuating fields or ropes\cite{biroRopes} 
are therefore introduced.  The ropes result from the geometrical overlap 
of wounded nucleons.  In the region of the overlapping wounded nucleons, 
a larger energy-density is felt during their fragmentation. 
This leads to an increase in both $s - \bar{s}$ pair production 
and $qq - \bar{q}\bar{q}$ pair production.
This mechanism naturally enhances strange meson and
multistrange baryon production.   These mechanisms will now 
be reviewed in more detail.

\section{Baryon Junction Stopping and Junction Loop Pair Production}
The baryon junction was first proposed in the 70's in order to understand 
baryon - antibaryon annihilation\cite{rossi77}.  Recently, 
the baryon junction was resurrected to provide an efficient mechanism
for baryon stopping\cite{khar96}.  

The baryon junction is motivated from writing the QCD gauge 
invariant operator of the baryon.  It is the 
vertex which links the three color flux (Wilson) lines flowing 
from the valence quarks.   Since the junction is a gluonic configuration,  
it can be easily transported into the mid-rapidity 
region in hadronic reactions.   At mid-rapidity, a baryon which is composed
of three sea quarks is produced from the fragmentation of the strings 
which connect the junction to the three valence quarks.
This gluonic mechanism is able to explain\cite{kopHera} a striking, 
preliminary measured baryon asymmetry observed approximately 8 units 
of rapidity away from the proton's fragmentation region in $ep$ 
collisions at HERA.

Using this Regge motivated mechanism, event generator calculations were 
able to reproduce the baryon stopping at the SPS\cite{vanceHijb}.  
However, since this baryon stopping mechanism depleted the 
available phase space in the central rapidity region for antibaryon 
production, a mechanism for antibaryon production was needed. 
The valence baryon junction mechanism was then extended and  
a new mechanism for antibaryon production was proposed\cite{vanceHijbb}.   
Like the valence baryon junction mechanism, this junction-anti-junction 
loop $(J\bar{J})$ mechanism is also derived from the topological gluon 
structure of the baryon and originates in the context of Regge phenomenology.
The $(J\bar{J})$ mechanism was shown to strongly enhance the anti-hyperon 
yields in nuclear collisions and 
to provide reasonable anti-hyperon to hyperon ratios.
Both of these processes were implemented in HIJING/B\=B,
a modified version of the HIJING event generator.

\begin{figure}
\begin{center}
\epsfxsize=5.0in   
\epsfysize=5.0in
\epsfbox{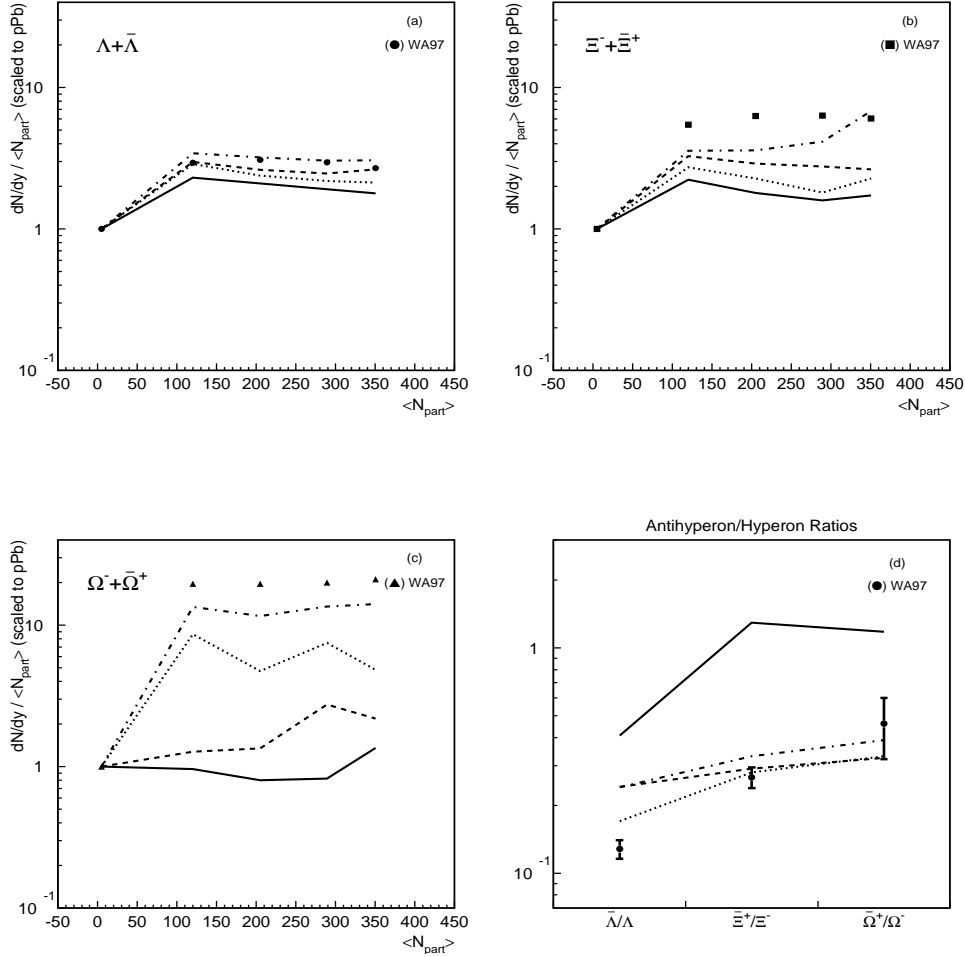}
\caption{The Hyperon yields per number of participant for $PbPb$ collisions
divided by the hyperon yields per number of participant in $pPb$ are plotted
in parts (a-c).  The ratios of the anti-hyperons to hyperons are shown in 
part (d).   Calculations from HIJING (solid), 
HIJING/B\=B (dashed), HIJING/B\=B+GCP (dotted) and 
HIJING/B\=B+GCP+Ropes (dash-dotted) are compared with measurements from 
WA97 Collaboration \protect \cite{wa97_qm99, wa97_plb}  }
\label{wa97_yields}
\end{center}
\end{figure}

The ratio of hyperon yields per number of participants in $PbPb$ 
collisions to the hyperon yields per number of participants in $pPb$ 
are plotted for $\Lambda$, $\Xi$ and $\Omega$ production in 
figures~\ref{wa97_yields}a-c. 
Comparing the calculations of HIJING (solid line) 
and HIJING/B\=B (dashed line) shows that the junction processes 
in HIJING/B\=B provide a slight enhancement between the $pPb$ and 
$PbPb$ systems.  However, it should be noted that the 
junctions have already strongly enhanced the $\Omega$ production
in $pPb$ by a factor of 10.  The antihyperon to hyperon ratios
are shown in figure~\ref{wa97_yields}d, where the junction mechanisms
are shown to provide more reasonable values.

\section{Final State Interactions}
In collisions between large nuclei, the hadrons which are 
produced from direct production mechanisms
such as string fragmentation are able interact with one another 
as they flow outward. 
At SPS energies, these interactions have typically 
low center of mass energies ($\sqrt{s} \le 3 \;{\rm GeV}$)       
and have been shown to influence the observed particle 
yields\cite{venus,rqmd,urqmd,dpmBab}.
The effects of the final states interactions, in particular the strangeness 
changing processes, are explored by coupling 
the Genera Cascade Program (GCP) \cite{pangGcp} to the HIJING/B\=B model.

The GCP \cite{pangGcp} models the evolution of
the produced hadrons by solving the relativistic Boltzmann equation with
two body collisions;
\beqar
p^\mu \partial_\mu W_a(\vec{x},\vec{p},t)  \nonumber \\
	 = \sum_{b_1,b_2} \int 
\left [ \prod_{i=1,2} \frac{d^3 \vec{p}_{b_i}}{(2\pi)^3 2 E_{b_i}} 
W_{b_i}(\vec{x},\vec{p},t) \right ] |\vec{v}_1 - \vec{v}_2 | 
\sigma_{b_1 + b_2} \nonumber \\
\times 
\left [ - \delta_{ab_1} \delta^3(\vec{p} - \vec{p}_{b_1})
 - \delta_{ab_2} \delta^3(\vec{p} - \vec{p}_{b_2})  
+  \sum_{j=1}^{m}\delta_{ac_j} \delta^3(\vec{p} - \vec{p}_{c_j}) \right ],
\nonumber \\ 
\eeqar{rbm4}
where $W_a$ is the particle phase space distribution and $\sigma_{b_1,b_2}$
is the cross section for the interaction of particles $b_1$ and $b_2$. 
A point particle distribution is assumed for the $W_a$, 
\beq
W_a(\vec{x},\vec{p},t) = \frac{N_a}{M_a} \sum_{i=1}^{M_a} 
\delta^3(\vec{x} - \vec{x}_i(t))
\delta^3(\vec{p} - \vec{p}_i(t)),
\eeq{rbm3}
where $M_a$ is the number of test partons ($M_a = 1$ is assumed for 
the present calculation).

In the GCP simulation, collisions between the hadrons occur 
when their distance of closest approach, $d$, is less than their 
interaction range $r = \sqrt{\sigma/\pi}$, which is determined from their 
interaction cross section, $\sigma$.    
As the system evolves, all possible future collisions are 
time ordered in the global center of mass frame.  
As a collision occurs, the branching ratios of the possible channels 
give the final state spectrum and all future possible collisions in 
the system are again determined.  Between collisions, the particles 
are assumed to travel along straight lines.  
The cross sections for the interactions are determined by fitting 
available data or are assumed from general arguments.  The inverse of 
these processes is obtained by modifying the cross section 
with the appropriate spin and momentum dependent factors 
(see reference~\cite{urqmd}).

When coupling GCP to HIJING/B\=B\cite{vanceHijGcp}, 
the phase space distribution of the hadrons produced from the string 
fragmentation is obtained
using the strong assumption that the pseudo-rapidity 
$\eta = 1/2 \ln [(t+z)/(t-z)] $ equals the rapidity 
$y = 1/2 \ln [(E+p_z)/(E-p_z)]$, or $\eta = y$.   

\begin{figure}
\begin{center}
\epsfxsize=5.0in   
\epsfysize=2.5in
\epsfbox{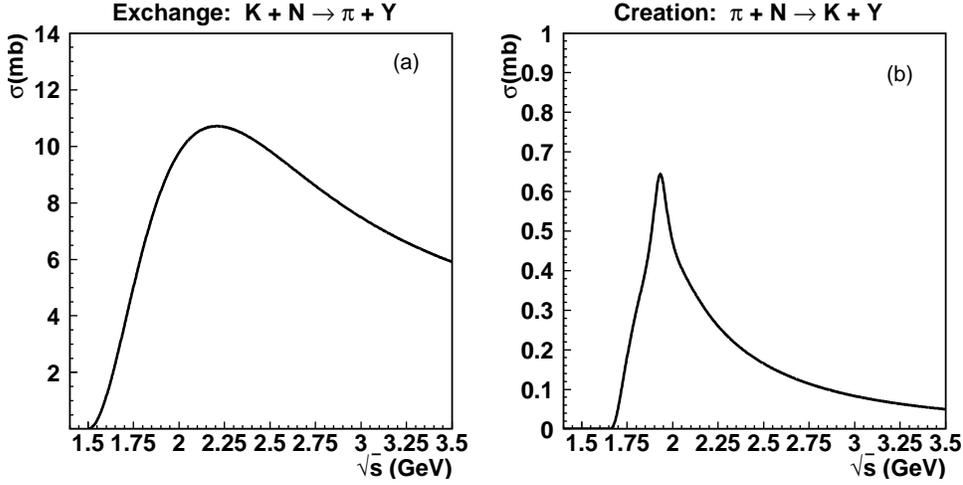}
\end{center}
\caption{The general strangeness exchange cross section and the general
strangeness creation cross section are shown in parts (a) and (b),
respectively.}
\label{sec_fits}
\end{figure}

Three categories of final state interactions influence 
hyperon yields.  The first is the strangeness exchange interactions, 
\beqar
N + K &\leftrightarrow& \Lambda + \pi,  \; \;  
N + K \leftrightarrow \Sigma + \pi, \nonumber \\
\Lambda + K &\leftrightarrow& \Xi + \pi, \; \;
\Sigma + K \leftrightarrow \Xi + \pi, \nonumber \\
\Xi + K &\leftrightarrow& \Omega + \pi .  
\eeqar{fs10}
Channels which include the $K^*$ and $\Delta$ resonances are also 
included.   For the $K^- + p$ and $K^- + n$ interactions, the cross
sections are obtained by fitting the data\cite{baldini88}. 
For the other reactions, a general strangeness exchange cross section,
$\sigma_{exchange}$ = $\sigma^{in}_{K^- + p} 
- \sigma^{in}_{K^+ + p} - \sigma_{K^- + p \rightarrow \bar{K}^0 + n}$ 
is used (see figure~\ref{sec_fits}a).
The strangeness exchange interactions are exothermic and are efficient
mechanisms for enhancing the hyperons.  The most effective of these 
reactions is the $K^*$ interacting with the $N$, $\Lambda$ and $\Xi$.

The second set of interactions are the strangeness creation reactions,
\beqar
N + \pi &\leftrightarrow& \Lambda + K, \; \;
N + \pi \leftrightarrow \Sigma + K, \nonumber \\
\Lambda + \pi &\leftrightarrow& \Xi + K, \; \;
\Sigma + \pi \leftrightarrow \Xi + K, \nonumber \\
\Xi + \pi &\leftrightarrow& \Omega + K. 
\eeqar{fs11}
The cross sections for the  $\pi^+ + p$ and $\pi^- + p$ 
interactions are obtained from fitting data\cite{baldini88} 
while the cross section used for the other channels is assumed 
to be $\sigma_{creation} 
= \sigma_{\pi^+ p \rightarrow \Sigma^+ K^+}$ 
(see figure~\ref{sec_fits}b).
Unlike the strong enhancements found in 
reference \cite{dpmBab}, the strangeness creation 
interactions in our simulation appears to have little effect 
in enhancing hyperon production. 

The third type of interactions is baryon-antibaryon annihilation
\beq
B + \bar{B} \rightarrow X,
\eeq{ann}
which depletes both the net baryon and antibaryon yields. 
The cross section used for this interaction is taken from 
reference~\cite{koch89}.

\begin{figure}
\begin{center}
\epsfxsize=3.5in   
\epsfysize=3.5in
\epsfbox{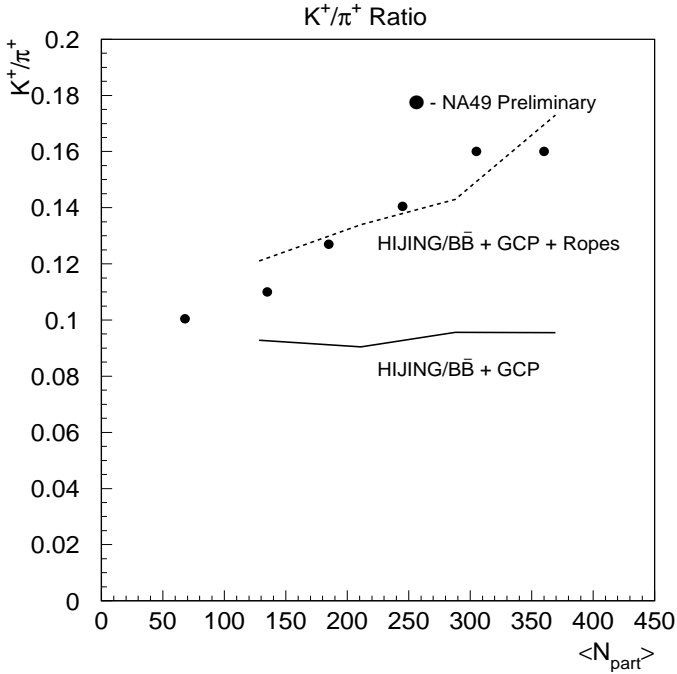}
\caption{Calculations of the $K^+/\pi^+$ ratio as a function of the number
of participants from HIJING/B\=B+GCP 
(solid line) and HIJING/B\=B+GCP+Ropes (dashed line) are compared 
with preliminary NA49 measurements\protect \cite{na49_qm99} 
for $PbPb$ collisions at the SPS.}
\label{ktopi}
\end{center}
\end{figure}

Calculations of HIJING/B\=B+GCP (dotted line) are shown in 
figure~\ref{wa97_yields}.  The final state strangeness exchange 
interactions dramatically enhance the $\Omega$.   However, the 
baryon-antibaryon annihilation decrease the total yields.
More striking 
is the comparison between HIJING/B\=B + GCP calculations of the  
$K^+/ \pi^+ $ ratio (ratio of the total $4\pi$ yields) 
with the preliminary data\cite{na49_qm99} 
as shown in figure~\ref{ktopi}.  Here, HIJING/B\=B + GCP gives
a dependence of this ratio on the number of participants which is flat,  
underestimating by a factor of 2 the value 
in the most central collisions.  
The junction and final state interactions
mechanisms most strongly influence hyperon production while only slightly 
enhancing the $K$ yields.
Additional strangeness enhancements mechanisms are therefore needed.

\section{Ropes or Transient Field Fluctuations}
Another efficient strangeness production mechanism is transient field
fluctuations or ropes. Ropes occur when wounded nucleons 
physically overlap during their fragmentation\cite{biroRopes}.  
When the wounded nucleons overlap, the chromo electric field strength 
or string tension, $\kappa$, is increased, 
allowing for enhanced production 
of $s-\bar{s}$ and $qq-\bar{q}\bar{q}$ pairs during fragmentation.  
For example, using the Schwinger particle 
production mechanism\cite{schwinger51}, the probability for producing 
an $s-\bar{s}$ pair to a $u-\bar{u}$ pair is given as 
\beq
P_{s/u} = e^{- \pi (m_s^2 - m_u^2) / \kappa }.
\eeq{schwinger}
In $pp$ interactions, $\kappa \simeq 1 \; {\rm GeV/fm} = 0.2 \; {\rm GeV^2}$,
and when using the constituent masses for the quarks 
(i.e. $m_s = 0.45 \; {\rm GeV}$ and $m_u = 0.325 \; {\rm GeV} $),
$P_{s/u} \sim 0.3$.   As $\kappa$ increases, the $P_{s/u}$ probability 
also increases and thus more strange hadrons are produced.  
This mechanism has been shown to increase hyperon production
\cite{rqmd,urqmdRopes}.  

When adding the ropes to HIJING/B\=B+GCP, the string tension grows
as the square-root of the number of overlapping wounded nucleons, 
$\kappa = \sqrt{n} \; {\rm GeV/fm} $.  
The number of overlapping wounded nucleons is determined by counting those
which lie within a radius of $r = 0.25 \; {\rm fm}$. 
The addition of ropes to HIJING/B\=B+GCP will be denoted 
by HIJING/B\=B+GCP+Ropes.

Using HIJING/B\=B+GCP+Ropes, calculations of 
the hyperon yields and the $K^+/\pi^+$ ratios are shown in 
figures~\ref{wa97_yields} and ~\ref{ktopi}, respectively.  As seen in 
figure~\ref{wa97_yields}a-c, the ropes enhance the hyperon production.  
As seen in figure~\ref{ktopi}, the ropes also reproduce the 
smooth increase of the $K^+/\pi^+$ ratio with the 
increased number of participants.  

\section{Conclusion}
The junction baryon stopping and junction loop baryon pair production 
mechanisms are important in describing the baryon stopping,
hyperon production and the ratio of antihyperons to hyperons at the SPS. 
In addition, final state interactions,
 whose effects are most dominantly observed through the 
$p_T$ distributions of the particles, are also important in changing 
the hyperon yields.  Here, the most dominant processes
are the strangeness exchange channel and baryon-antibaryon annihilation.
However, even with these two mechanisms, the 
overall net strangeness enhancement is under-produced as seen most strikingly 
through the $K^+/ \pi^+$ ratio. 
Another mechanism is therefore needed and the initial state rope mechanism 
is used.  This rope mechanism depends upon the number of overlapping
strings and thus is able to produce more strange particles in 
denser systems as
consistent with the observed $K^+/ \pi^+$ ratio. 
When combined, the three mechanisms are able to describe both the 
large strangeness enhancement in the 
dense $PbPb$ collisions and the smooth dependence of the enhancement 
on the number of participants as seen at the SPS.

\section{Acknowledgments}
I would like to thank M. Gyulassy, T. Biro and N. Xu for helpful 
discussions.  This manuscript was authored under Contract No. 
DE-AC02-98CH10886 with the U. S. Department of Energy.

\end{document}